 \documentclass[preprint,review,12pt]{elsarticle}

\usepackage{epsfig}
\usepackage{amsmath}
\usepackage{amssymb}
\usepackage{mathptmx}
\usepackage{bm}
\usepackage{verbatim}
\usepackage{epsfig}
\usepackage{graphicx}
\usepackage{hyperref}					
\hypersetup{colorlinks,
	linkcolor=blue,%
	citecolor=blue}

\graphicspath{{./figures/}}
\usepackage[section]{placeins}
\usepackage{caption}
\usepackage{booktabs}  
\usepackage{threeparttable} 
 \usepackage{amsthm}
\usepackage[left=2.5cm, right=2.5cm, lines=45, top=0.8in, bottom=1.0in]{geometry}
\usepackage{blindtext}
\journal{JMS}

\begin{document}
\begin{sloppypar}
\begin{frontmatter}


\author[mymainaddress]{Liliang Tian}
\author[mymainaddress]{Humin Duan}
\author[mymainaddress]{Jiaming Luo}
\author[mymainaddress]{Yonghong Cheng}
\author[mymainaddress]{Le Shi\corref{mycorrespondingauthor}}
\cortext[mycorrespondingauthor]{Corresponding author}

 \address[mymainaddress]{State key Laboratory of Electrical Insulation and Power Equipment, Centre of Nanomaterials for Renewable Energy, School of Electrical Engineering, Xi’an Jiaotong University, Xi’an 710049, China.}
 \ead{le.shi@mail.xjtu.edu.cn}

\title{A superior two-dimensional nanoporous graphene membrane for hydrogen separation}



\begin{abstract}
Hydrogen is one of the prime candidates for clean energy source with high energy density. However, current industrial methods of hydrogen production are difficult to provide hydrogen with high purity, thus hard to meet the requirements in many application scenarios. Consequently, the development of large-scale and low-cost hydrogen separation technology is urgently needed. In this work, the gas separation properties of a newly synthesized two-dimensional nanoporous graphene (NPG) membrane material with patterned dumbbell-shape nanopores are investigated. The permeation energy barriers of different gases through this membrane are calculated using the density functional theory (DFT) calculations. Molecular dynamics (MD) simulations are also employed to study the permeation behavior of H${_2}$ in binary mixtures with O${_2}$, CO${_2}$, CO, and CH${_4}$. Both the DFT and MD calculation results show that this newly synthesized NPG membrane material can provide a high permeability as well as an ultrahigh selectivity simultaneously, making it a prospective H${_2}$ separation membrane with superior performance.

\end{abstract}


\begin{keyword}
\texttt {$\rm {H_2}$ purification, Nanoporous graphene (NPG), 2D membrane, Density functional theory (DFT), MD simulation.}

\end{keyword}




\end{frontmatter}


\section{Introduction}
Hydrogen is one of the prime candidates which may replace fossil energy sources for its cleanliness, low cost, environmental friendliness and high energy density \cite{OsterlohInorganicnanostructuresphotoelectrochemical2013,WangRecentProgressCobaltBased2016}. It also has a wide range of applications in areas such as fuel cells \cite{EdwardsHydrogenfuelcells2008}, electronics industry \cite{OkolieFuturisticapplicationshydrogen2021} and pharmaceutical synthesis \cite{MurugesanCatalyticreductiveaminations2020}. However, the well-established industrial hydrogen production methods produce a variety of gaseous byproducts, such as carbon monoxide, carbon dioxide and methane from methane steam reforming, and nitrogen from ammonia decomposition \cite{LangReviewHydrogenProduction2011}. These impurities can have serious adverse impacts on the application of hydrogen, such as the toxic effects of carbon monoxide on catalysts and the permanent damage to fuel cells caused by carbon dioxide and methane \cite{AhluwaliaEffectCOCO22008}. Therefore, hydrogen separation technology with low cost is urgently needed to prepare high-purity hydrogen.

The mainstream technologies for industrial hydrogen separation including variable pressure adsorption, cryogenic separation and membrane separation, among which membrane separation has been expected to be less mechanically complex, less energy demanding and less costly in terms of capital \cite{DolanNonPdBCCalloy2010,AlimovHydrogentransporttubular2018,Luopportunitymembranetechnology2020}. Unfortunately, both the well-developed polymeric membranes and the recently reported materials like metal-organic frameworks are unable to offer a satisfactory performance due to the well-known tradeoff between selectivity and permeability \cite{Robesonupperboundrevisited2008a}. The emergence of single-atom-thick graphene provide a possible solution to break this tradeoff, since the atomic membranes provide the shortest diffusion paths for molecules, while the crystal structures with highly ordered pore arrays guarantee high selectivity. However, due to the high electron density of aromatic rings, intact graphene and graphene with small localized defects are impermeable to any gases \cite{BunchImpermeableAtomicMembranes2008,LeenaertsGrapheneperfectnanoballoon2008}. Therefore, the aim of high permeability can only be achieved by graphene with nanopores. In 2012, Koenig et al. \cite{KoenigSelectivemolecularsieving2012} firstly demonstrated that the sub-nanometer pores on graphene formed by UV-induced etching can achieve selective transportation of gases, and the experimental results qualitatively agree with the theoretical predictions made by Blankenburg et al. \cite{BlankenburgPorousGrapheneAtmospheric2010}. In 2014, Celebi et al. \cite{CelebiUltimatePermeationAtomically2014a} fabricated a millimeter-sized graphene membrane with controlled pore sizes which can provide ultrahigh H${_2}$ permeability.

Although the hydrogen permeability of two-dimensional (2D) porous graphene membranes are several orders of magnitude higher than that of pristine graphene, their selectivity is still limited \cite{CelebiUltimatePermeationAtomically2014a,WangFundamentaltransportmechanisms2017b}. The kinetic diameter of most gases are in the range from 2.5 Å to 4 Å, which are very close to each other. Therefore, achieving high selectivity requires pore size modulation at sub-nanometer or even smaller scale \cite{YuanAnalyticalPredictionGas2019a}. Previous studies have reported a variety of possible solutions to tune the selectivity, such as changing the shape of the nanopores \cite{RaghavanH2CH4Gas2017}, replacing the atoms at the pore edges \cite{LiuInsightsCO2N22013,DaryabariHeliumselectivitydoped2020}, modulating the pore size by stress and so on \cite{ZhuTheoreticalstudytunable2016}. These reports theoretically predicted possible ways to achieve high selectivity, but it is very challenging to realize these ideas with the current preparation technologies. So far, the preparation technologies of nanoporous graphene fall into two categories: 1) top-down, where nanopores are created on intact graphene by oxidative etching or ion bombardment; and 2) bottom-up, where porous graphene with certain nanopore structure is chemically synthesized \cite{Novoselovroadmapgraphene2012}. Bottom-up synthesis of nanoporous graphene has a clear advantage on pore size modulation over the top-down preparation methods, which cannot control pore morphology as well as size in a large scale and precise manner. Several bottom-up synthesized nanoporous 2D materials have demonstrated high selectivity for specific gases. For example, graphdiyne \cite{CranfordSelectivehydrogenpurification2012}, a carbon allotrope with patterned triangular nanopores, can effectively separate hydrogen from synthesis gases containing carbon monoxide and methane; and inorganic graphenylene \cite{WangHighefficiencyheliumseparation2020}, a boron nitride analogue of grapheylene, separates helium from natural gas. In addition, a number of carbon-based nanoporous 2D materials have been experimentally synthesized recently, and their properties for gas separation need to be further explored via either experimental or theoretical approaches \cite{BrockwayNobleGasSeparation2013,MorenoBottomupsynthesismultifunctional2018,LiSynthesisggraphynemechanochemistry2018,LiGraphdiynegraphynetheoretical2014}.

\begin{figure*}[h]
   \centering
   \includegraphics{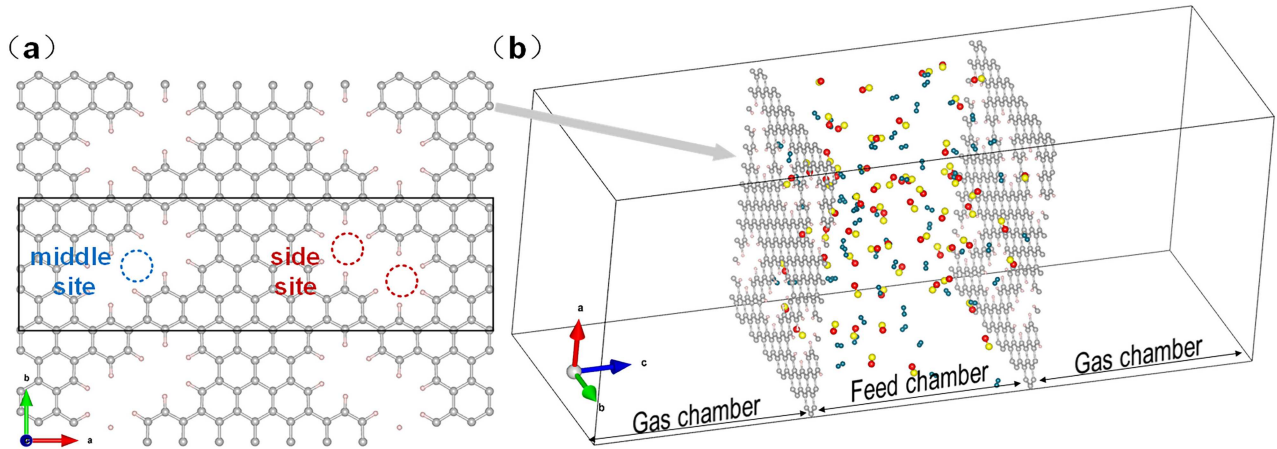}
   \caption{(a) Optimized NPG cell (b) Model system used in molecular dynamics simulations, with O atoms in red, C (NPG) atoms in gray, C (CO) atoms in yellow, H (H${_2}$) atoms in blue and H (NPG) atoms in pink. In all model diagrams of this paper, the C/H atoms of the gas and the C/H atoms of the NPG are distinguished by different colors.}
   \label{fig:1}
\end{figure*}

In 2018, César Moreno et al. \cite{MorenoBottomupsynthesismultifunctional2018} reported that graphene nanoribbons (GNR) designed with atomic-level precision can went through a dehydrogenation cross-coupling reaction and form nanoporous graphene (NPG) with high surface area, high pore density and inherently well-defined long-range ordered nanopores. The geometry of nanopores in NPG is quite unique as shown in Figure \ref{fig:1}(a), which shows a dumbbell shape with the C atoms at the edge of the pore passivated by H atoms. The dumbbell-shaped nanopore may restrict the passage of gases with large kinetic diameter while provide abundant penetration routes for gas molecules with small kinetic diameter, thus achieving a high permeability and selectivity simultaneously. In this work, we investigated the adsorption properties of common light gases on NPG by means of first principle calculations, and then further calculated the energy barriers for the permeation of these gases through NPG to estimate their permeability. Molecular dynamics simulations were employed to investigate the permeation behavior of H${_2}$ in mixtures with O${_2}$, CO${_2}$, CO and CH${_4}$, and to investigate the mechanisms affecting the permeation rate of H${_2}$. Our results show that this newly synthesized NPG can provide both high permeability and high selectivity for H${_2}$ when serving as a H${_2}$ separation membrane. 

\section{Method and computational details} 
\subsection{Density functional theory (DFT) calculations}
First principles calculations are performed using the ABINIT software package \cite{GonzebriefintroductionABINIT2005}. Projector-augmented-wave (PAW) potentials \cite{BlochlProjectoraugmentedwavemethod1994,Kresseultrasoftpseudopotentialsprojector1999b} were used to describe the electron-ion interactions, while the electron exchange-correlation interactions were treated using generalized gradient approximation (GGA) in the scheme of Perdew-Burke-Ernzerhof (PBE) \cite{PerdewGeneraLizedGradientApproximation1996}. The Broyden-Fletcher-Goldfarb-Shannon (BFGS) algorithm was used to optimize the structure of NPG and gas molecules with a maximum absolute force tolerance of 5$\times$10$^{-4}$ hartree/Bohr. Convergence criterion was set to be 5$\times$10$^{-5}$ hartree/Bohr for each SCF iteration. Grimme's DFT-D2 \cite{GrimmeSemiempiricalGGAtypedensity2006} corrections were adopted to account for the van der Waals interaction. The Monkhorst–Pack k-point mesh was set to be \textless 0.05 Å$^{-1}$ and the cut-off energy was set to be 20 Ha. A vacuum space of 2 nm in z-direction was added to avoid interactions between periodic images. Hirshfeld charge analysis was performed to analyze the charge distribution. Climbing Image Nudged Elastic Band (CINEB) algorithm\cite{HenkelmanImprovedtangentestimate2000,HenkelmancLimbingimagenudged2000} was used to search for the transition state of gas molecules passing through the NPG. The energy difference between the transition state and the adsorbed state was defined as the diffusion energy barrier.

\subsection{Molecular dynamics (MD) simulation}
The behavior of gas permeation through NPG was further investigated using MD simulations. The molecular dynamics simulations were carried out using LAMMPS software package \cite{PlimptonLAMMPSlargescaleatomicmolecular2007}. The size of the simulated system is 3.24 $\times$ 2.57$\times$12$\ \rm{nm^3}$. Two layers of NPG are placed in parallel to divide the box into three equal parts as shown in Figure 1b, as a result the volume ratio between the gas chamber space on both sides and the feed chamber in the middle region is 2:1. Periodic boundary conditions were applied in all three directions. 60 H${_2}$ molecules and 60 foreign gas molecules (i.e. CO, CO${_2}$,etc) were placed randomly in the feed chamber space before the simulation starts. The initial pressure of the feed chamber is 7.46 Mpa. In order to prevent the two layers of NPG from moving apart or shifting laterally during the simulation, a flexible NPG model with four corner carbon atoms fixed was used. A rigid NPG model with all C and H atoms fixed was tested for comparison (as elaborated in the Supporting Information). 

The C-C and C-H bonding interactions in the NPG were characterized using AMBER force field \cite{CornellSecondGenerationForce1995}. Interactions among gas molecules were characterized using an all-atom model, meanwhile the bond stretch and bond angle deformation in gas molecules were described by harmonic potentials (H${_2}$, CO${_2}$, CH${_4}$), morse potential (O${_2}$) and COMPASS potential (CO). The interactions between NPG and H${_2}$ were described by the AIREBO potential function \cite{Stuartreactivepotentialhydrocarbons2000a}, given its unique advantage in describing the interaction between C and H atoms. The interaction between the other atoms were all described by coupling the Lennard-Jones potential with a Coulomb term to account for van der Waals forces and short-range electrostatic interactions. Interaction parameters between crossed atoms were obtained by using the Lorentz-Berthelot mixing rule. All atoms on the NPG carry the charges obtained from the above DFT calculations to refine the interaction between the NPG and gases. Specific force field parameters can be found in the Supporting Information. All simulations were carried out with a timestep of 0.1 fs and a total simulation time of 20 ns. Trajectories during the MD simulation were recorded every 20,000 steps. The system was simulated in NVT ensemble with a Nosé–Hoover thermostat to maintain the system temperature at 300 K \cite{HooverCanonicaldynamicsEquilibrium1985}.

\section{Results and discussion}
\subsection{Gas adsorption on NPG}
The geometry of NPG was constructed based on the experimental reports \cite{MorenoBottomupsynthesismultifunctional2018} and optimized using DFT clauclation. The optimized NPG structure is shown in Figure 1 with lattice parameters of 32.4$\times$8.57 $\rm {Å^{2}}$. The size of the pores is approximately: 8.06$\times$3.11 ${\rm Å^{2}}$. According to the theory proposed by Drahushuk et al. \cite{DrahushukMechanismsGasPermeation2012}, the gas molecules are firstly adsorbed by the graphene onto its surface region close to the pores, then overcome the diffusion resistance and penetrate across the pores, and finally desorb from the graphene slab. Sun et al. \cite{SunMolecularsievinggraphene2017} suggested that although high porosity could increase the permeability, a smaller membrane surface area would result in a lower adsorption energy towards the gases, thus reducing the adsorption flux. Therefore, there may exists a tradeoff between the area of surface and pore. Based on the optimized structure of NPG, the ratio of surface area to pore area is approximately 9:2 and the pore density reaches 7.18$\times$10$^{17}$m$^{-2}$. Such a large proportion of pore area and the introduced hydrogen atoms may result in a different gas adsorption behavior compared with pristine graphene. In this work, the gas adsorption energy is defined as:
\begin{equation}
\label{1}
E_{\rm ads}=E_{\rm NPG}+E_{\rm Gas}-E_{\rm{NPG+Gas}}
\end{equation}
where $E_{\rm{NPG+Gas}}$ is the total energy of NPG adsorbed with gas molecule, and $E_{\rm{NPG}}$, $E_{\rm{Gas}}$ are the total energy of NPG and gas molecule respectively. We studied the adsorption behavior of different gas molecules on the pore region as well as on various high symmetry points of the surface region. For diatomic and triatomic molecules, adsorption modes with gas molecules parallel and perpendicular to the NPG surfaces were considered. The highest adsorption energies in both regions are shown in Table \ref{tab:1}.

Based on the calculation results, the adsorption energies of the gas molecules on NPG are generally small. O${_2}$ has the highest adsorption energy on the pore region of NPG (0.37 eV), which is much higher than that on pure graphene, while its adsorption energy on the surface region of NPG is close to the case of pristine graphene \cite{KaurAdsorptionCOO22018}. This indicates that the introduction of H atoms at the pore edges enhances the van der Waals and Coulombic interaction between NPG and O${_2}$. NPG also shows some adsorption capability towards CO${_2}$ with an adsorption energy of 0.15 eV, which is similar to that of pristine graphene \cite{Cabrera-SanfelixAdsorptionReactivityCO22009}. The adsorption energies of N${_2}$, CH${_4}$, NH${_3}$ and CO on NPG are all relatively close to each other, which are around 0.1 eV. In contrast, the interaction between H${_2}$ and NPG is quite weak (40 meV). The relatively low adsorption energy allows H${_2}$ to diffuse rapidly and frequently over most of the space, and the permeation through the pores of NPG is dominated by direct fluxes.

\begin{table}[htbp]
  \centering
  \renewcommand\arraystretch{1.2}
  \caption{Maximum adsorption energy and corresponding adsorption height of gas molecules in the pore region and surface region of the NPG.}
    \begin{tabular}{p{2cm}p{2cm}p{2cm}p{2cm}}
    \toprule
     \multicolumn{1}{l}{Gas} & \multicolumn{1}{l}{$E_{\rm ads}(eV)$} & \multicolumn{1}{l}{D ($\mathring{\rm A}$)} & Region \\
    \midrule
    {H$_2$} & 0.041 & 2.521  & pore \\
          & 0.033 & 2.962  & surface \\
    \midrule
    {O${_2}$} & 0.371 & 2.248  & pore \\
          & 0.251 & 3.112  & surface \\
    \midrule
    {CO${_2}$} & 0.151 & 3.270  & pore \\
          & 0.138 & 3.351  & surface \\
    \midrule
    {CO} & 0.102 & 2.433  & pore \\
          & 0.100  & 3.358  & surface \\
    \midrule
    {N${_2}$} & 0.120  & 2.545  & pore \\
          & 0.09  &  3.30  & surface \\
    \midrule
    {NH${_3}$} & 0.045  & 4.536  & pore \\
          & 0.113  & 3.466  & surface \\
    \midrule
    {CH${_4}$} & 0.020  & 4.997  & pore \\
          & 0.117  & 3.276  & surface \\
    \midrule
    {He} & 0.014  & 3.040  & pore \\
          & 0.020  & 3.20  & surface \\
    \midrule
    {Ne} & 0.065  & 2.110  & pore \\
          & 0.049  & 3.203  & surface \\
    \midrule
    {Ar} & 0.096  & 2.529  & pore \\
          & 0.075  & 3.340  & surface \\
    \bottomrule
    \end{tabular}%
  \label{tab:1}%
\end{table}%

\subsection{Energy barriers of gas penetration across NPG}
\begin{figure*}[h]
   \centering
   \includegraphics{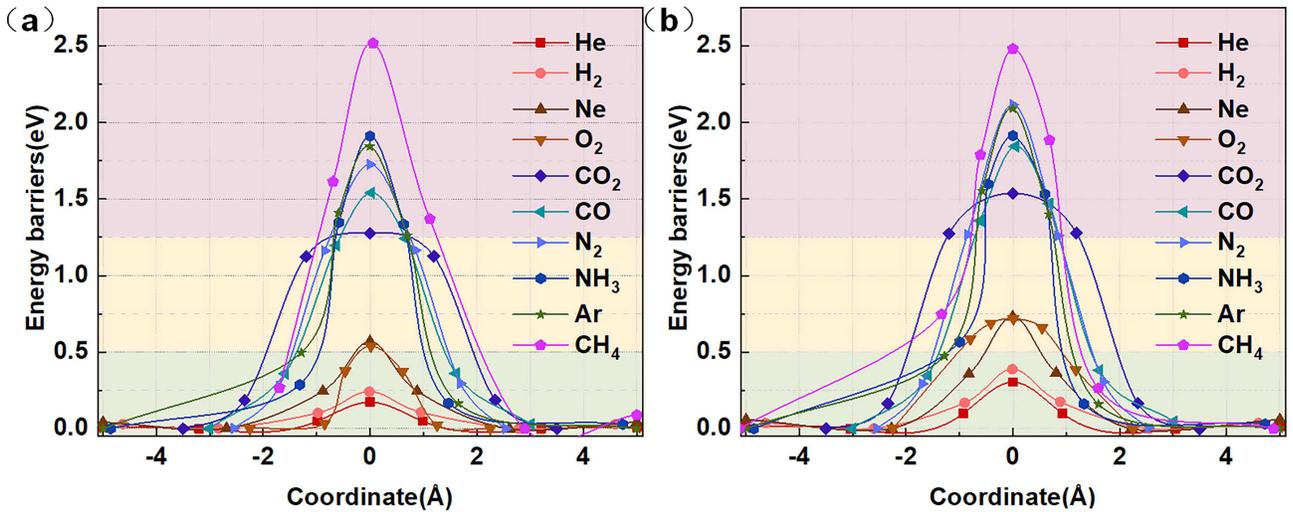}
   \caption{Energy barriers for gas permeation through the NPG at (a) the middle site and (b) the side site.}
   \label{fig:2}
\end{figure*}

In order to evaluate permeation capability of different gases, we calculated the permeation energy barrier of gases through NPG. Gas molecules may pass through the NPG from either the “side site” or the “middle site” of the pore as shown in Figure 1(a), so the energy barriers for each of the ten types of gas molecules through both sites were calculated, as shown in Figure \ref{fig:2}. It is found that except for O${_2}$ and NH${_3}$, where the energy barriers of permeation through both sites are similar, the energy barriers of the other gases are lower through the “middle site”, which means that the gas molecules can penetrate the NPG more easily through this site. The ten types of gas molecules can be roughly divided into three categories according to their permeation energy barriers through the “middle site”. The first group contains He and H${_2}$, which show low permeation energy barriers of 0.17 eV and 0.24 eV,respectively. This means that H${_2}$ and He can permeate through the NPG easily. The second category contains Ne and O${_2}$, which have moderate permeation energy barriers, both of which are around 0.55 eV. The last category contains the remaining six gases including CO${_2}$, CO, N${_2}$, Ar, NH${_3}$ and CH${_4}$, all of which have energy barriers above 1.5 eV and are difficult to permeate through NPG.

\begin{figure}[h]
   \centering
   \includegraphics{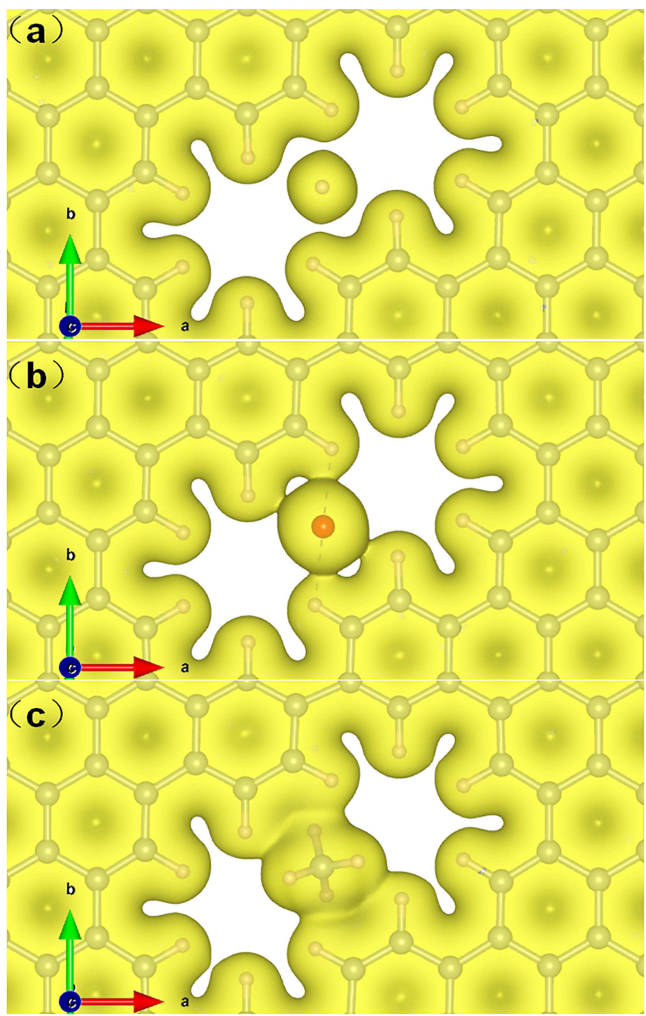}
   \caption{Electron density isosurfaces for (a) H${_2}$, (b) O${_2}$, (c) CH${_4}$ molecules passing through NPG. The value of isosurfaces is 0.02 e/Å$^{3}$.}
   \label{fig:3}
\end{figure}

We further analyzed the electron density isosurface of each gas in their transition state during the permeation process. As shown in Figture \ref{fig:3}, H${_2}$, O${_2}$ and CH${_4}$ were chosen to represent the situation of the three categories, and the value of isosurface was set to be 0.02 e/Å$^{3}$. H${_2}$ has a small molecular kinetic diameter and a low electron density, so it shows no electron overlap with the NPG, which agrees well with its low diffusion resistance. The kinetic diameter of O${_2}$ is 3.3 Å, which exceeds the length of the narrowest part of the pore. The O${_2}$ molecule has a larger possibility to penetrate NPG when it is perpendicular to the NPG slab, and the corresponding energy barrier is higher compared with the case of H${_2}$. The electron density of CH${_4}$ has a large overlap with that of NPG, making it difficult to realize penetration. The electron density isosurface for other gases in the transition state are shown in Figture S1.

The selectivity ($A$) of NPG towards H${_2}$ and He were calculated using the Arrhenius equation based on the above calculated permeation energy barriers. $A$ is defined as:
\begin{equation}
\label{2}
A_{X/Y}=\frac{r_X}{r_Y}=\dfrac{B_{X}e^{-E_{X}/RT}}{B_{Y}e^{-E_{Y}/RT}}    
\end{equation}
where $X$ and $Y$ represent the two gases, $r$ is the permeability, $B$ is the gas coefficient, $E$ is the permeation energy barrier, T is the temperature, and R is the molar gas constant. Here it is assumed that all gas coefficients are equal. The selectivity between different gases are listed in Table \ref{tab:2}. The calculations show that NPG is superior in the selective transportation of H${_2}$ and He, especially for H${_2}$. When calculating the energy barriers, we only considered the extreme case where the gas molecules are perpendicular to the NPG and did not take into account of other possible penetration modes, therefore the selectivity estimated can only provide a qualitative reference. 

\begin{table}[htbp]
  \centering
  \renewcommand\arraystretch{2}
  \caption{Selectivity of H${_2}$ as well as He for the permeation of other gases through NPG at 300k.}
    \begin{tabular}{p{2cm}p{2cm}|p{2cm}p{2cm}}
    \toprule
    $X/Y$   & \multicolumn{1}{l|}{$A(X/Y)$} & $X/Y$   & \multicolumn{1}{l}{$A(X/Y)$} \\
    \midrule
    H${_2}$/O${_2}$ & $1.0\times 10^8$ & He/Ne & $4.6\times 10^6$ \\
    H${_2}$/CO${_2}$ & $2.0\times 10^{17}$ & He/CO${_2}$ & $3.4\times 10^{18}$ \\
    H${_2}$/CO & $7.0\times 10^{21}$ & He/CO & $9.1\times 10^{22}$ \\
    H${_2}$/N${_2}$ & $7.0\times 10^{24}$ & He/N${_2}$ & $1.2\times 10^{26}$ \\
    H${_2}$/NH${_3}$ & $1.0\times 10^{28}$ & He/Ar & $1.1\times 10^{28}$ \\
    H${_2}$/CH${_4}$ & $4.0\times 10^{37}$ & He/CH${_4}$ & $2.3\times 10^{39}$ \\
    \bottomrule
    \end{tabular}%
  \label{tab:2}%
\end{table}%

\subsection{MD simulation of H${_2}$ separation}
\begin{figure*}[h]
   \centering
   \includegraphics{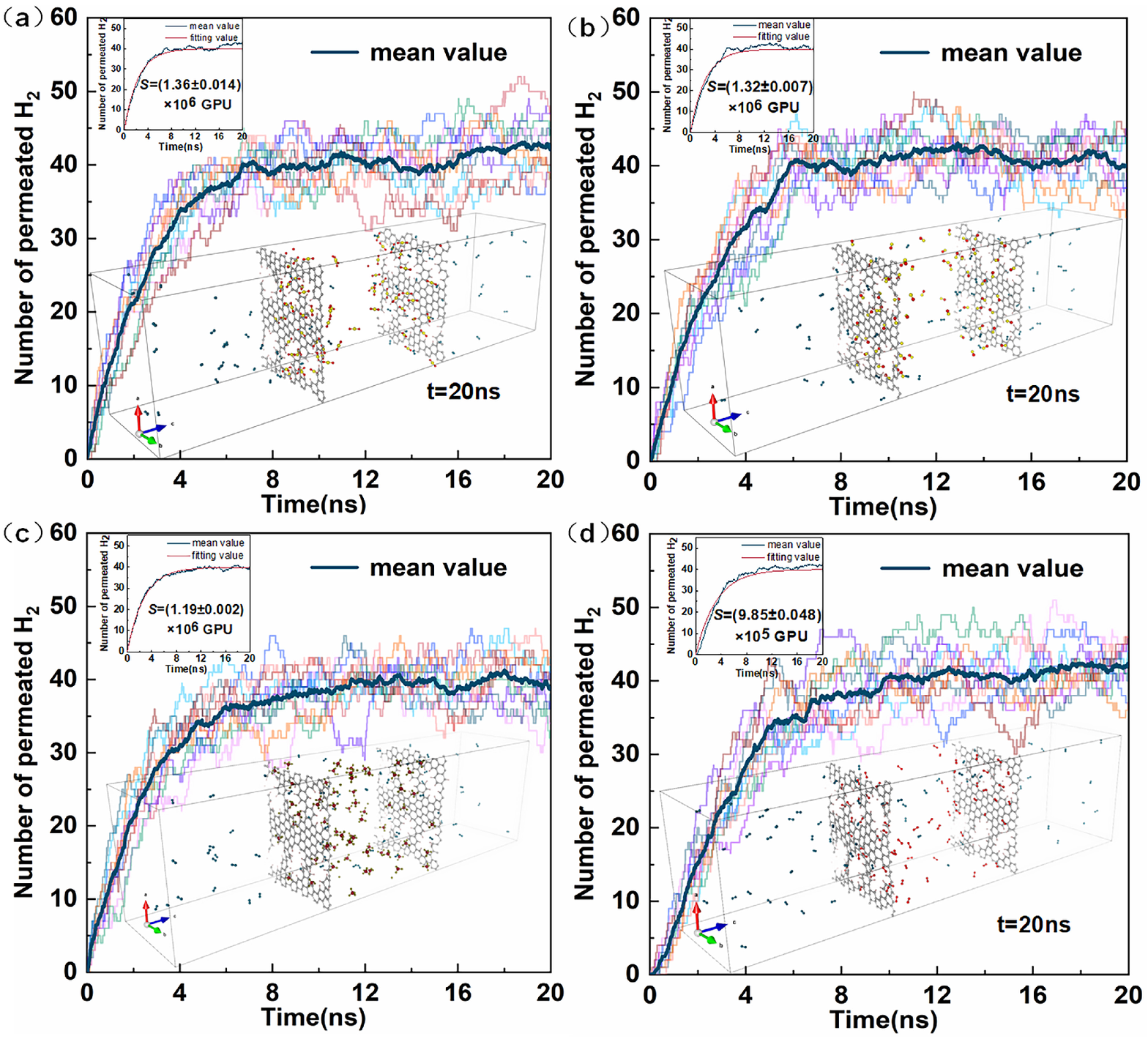}
   \caption{Molecular dynamics simulations of H${_2}$ permeation results for (a) H${_2}$/CO${_2}$, (b) H${_2}$/CO, (c) H${_2}$/CH${_4}$ (d) H${_2}$/O${_2}$ combinations. The slim rainbow coloured lines are the result of ten single runs, the thick dark blue line is the average of the results of all runs. The insets at the upper left corner show the average values and their corresponding fitted values and the $S$ in the inset is the permeability of H${_2}$ converted from the fitted results. The insets on the bottom side are snapshots of the simulated system at 20 ns.}
   \label{fig:4}
\end{figure*}

In order to further evaluate the performance of NPG towards H${_2}$ separation, MD simulation was employed to study the gas permeation behavior of four combinations of gases: H${_2}$/CO${_2}$, H${_2}$/CO, H${_2}$/CH${_4}$ and H${_2}$/O${_2}$. The simulation results showed that none of the foreign gas molecules (CO${_2}$, CO, CH${_4}$ and O${_2}$) permeated through the NPG after 20 ns MD simulation, and the NPG exhibited 100$\%$ selectivity towards H${_2}$. The variation of the number of H${_2}$ molecules permeated as a function of time is shown in Figure \ref{fig:4}. It can be found that H${_2}$ molecules will reach permeation equilibrium at around 5-6 ns, therefore the simulation time setting of 20 ns in this work is reasonable.
Apparently, the relation between the number of H${_2}$ molecules permeated and time is none linear. In order to calculate the permeability of H${_2}$ molecules precisely, the relationship between the permeability of H${_2}$ and time in the ideal situation is derived from the flux definition of the permeability $S$. The gas flux $J$ can be defined as:
\begin{equation}
\label{3}
J={\frac{dN}{d\tau}}{\frac{1}{\rm N_A}}=S\Delta P{\rm A_{NPG}}
\end{equation}
where $N$ is the number of permeable molecules, $\rm N_A$ is Avogadro’s constant, and $\rm A_{NPG}$ is the membrane area. $\Delta P$ is the pressure difference between the gas chamber and feed chamber, which is a function of the number of gas molecules on both sides of the membrane, and can be calculated as:
\begin{equation}
\label{4}
\Delta P=({\frac{60-N-N_{ad}}{60}-\frac{N}{60\times2}}){\rm P_{IN}}
\end{equation}
where $\rm P_{IN}$ is the initial pressure and $N_{ad}$ is the number of molecules adsorbed on the NPG surface, which can be obtained from simulations. It can then be deduced that $N$ as a function of time $\tau$:
\begin{equation}
\label{5}
\frac{dN}{d\tau}=9.35\times10^{11}S(40-N-\frac{2N_{ad}}{3})
\end{equation}
\begin{equation}
\label{6}
N=(40-\frac{2N_{ad}}{3})(1-e^{-9.35\times10^{11}S\tau})
\end{equation}
Based on Equation \ref{6}, we fit the MD simulation results of the four kinds of gas combinations. To increase the fitting accuracy, ten simulations were performed for each gas combination, and then the average of all simulated results was adopted in the fitting, as shown in the inset of Figure \ref{fig:4}. It can be found that our fitted curves are very close to the simulation results, and the R-Square of all the fitted results is greater than 0.975. The obtained fitting parameters show that H${_2}$ has a relatively close permeation rate of 1.36×10$^{6}$ GPU and 1.32×10$^{6}$ GPU in the combinations of H${_2}$/CO and H${_2}$/CO${_2}$, respectively. On the other hand, the H${_2}$ permeation rate in the H${_2}$/CH${_4}$ (1.19×10$^{6}$ GPU) and H${_2}$/O${_2}$ (9.85×10$^{5}$ GPU) is relatively lower. The H${_2}$ permeation rates in all cases are on the order of 10$^{6}$ GPU, far exceeding the H${_2}$ permeation rates of conventional polymer membranes. Therefore, it can be concluded that NPG accomplishes permeation rates up to 10$^{6}$ GPU while maintaining superior selectivity for H${_2}$. 

\begin{figure*}[h]
   \centering
   \includegraphics{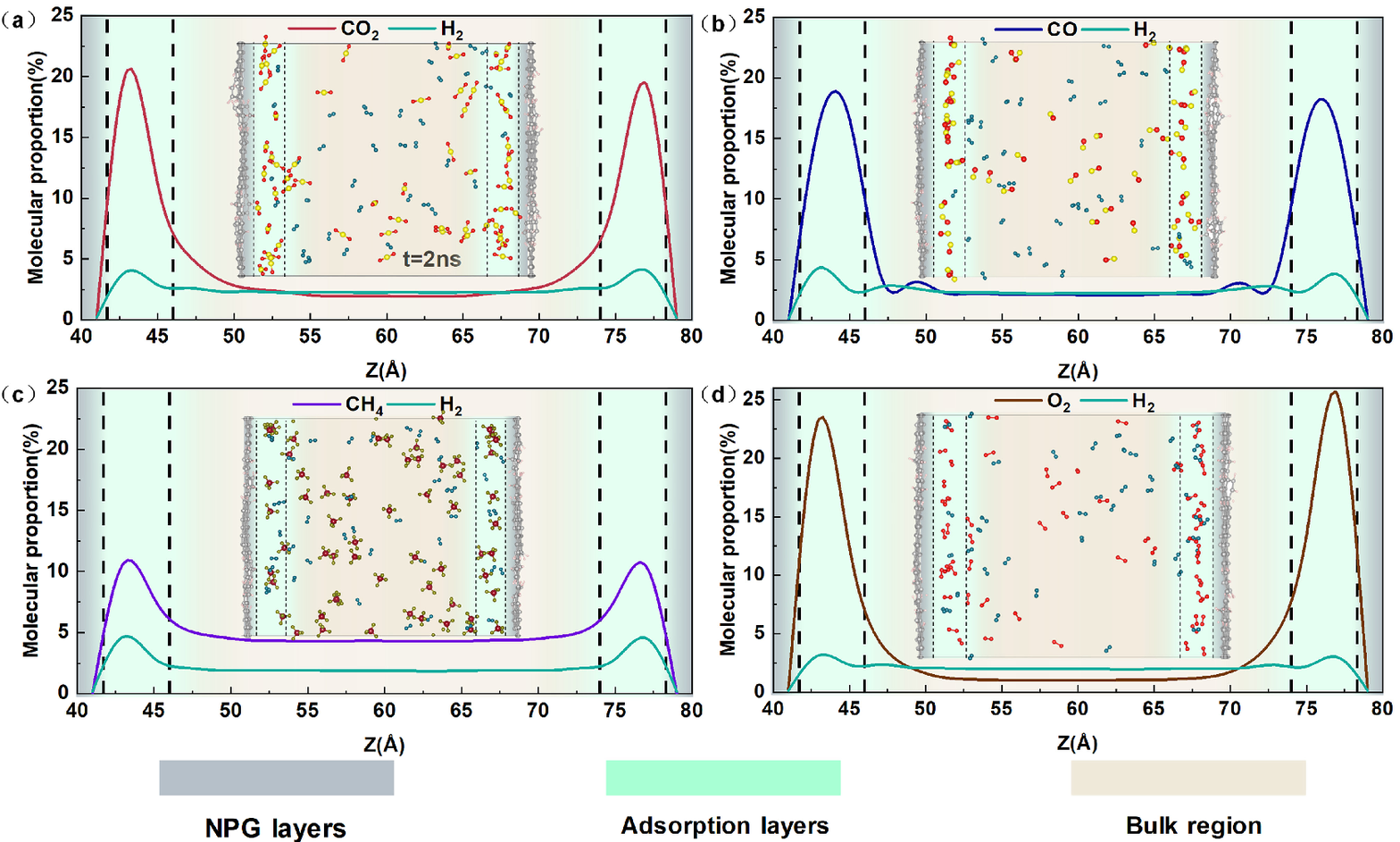}
   \caption{The proportional distribution of H${_2}$ molecules and foreign gas molecules in different gas molecule combinations along the $z$-direction in the feed chamber space.}
   \label{fig:5}
\end{figure*}

The lower H${_2}$ permeability in the case of H${_2}$/O${_2}$ can be attributed to the higher adsorption energy of O${_2}$ on NPG surface. We counted the gas distribution along the z-direction in the gas feed chamber and divided it into three different regions following the definition in the study of Sun et al. \cite{SunMechanismsMolecularPermeation2014} as shown in Figure \ref{fig:5}. The spaces of 4 nm \textless\ z \textless 4.17 nm and 7.83 nm \textless\ z  \textless 8 nm, where the NPG layers exist, are defined as the NPG region; the spaces of 4.17 nm \textless\ z \textless 4.6 nm and 7.4 nm \textless\ z \textless 7.83 nm, where the first layer of adsorbed molecules exist, are defined as the adsorption layer, and the space of 4.6 nm \textless\ z \textless 7.4 nm is defined as the bulk region. As shown in Figure 5, the ratio of O${_2}$, CO${_2}$, CO and CH${_4}$ molecules in the adsorption layer are 84$\%$, 71.7$\%$, 72$\%$ and 47$\%$, respectively. The corresponding ratio of H${_2}$ molecules in the adsorption layers of the four cases are 15.4$\%$, 19.1$\%$, 18.6$\%$ and 19.25$\%$, respectively. According to Sun et al \cite{SunMechanismsMolecularPermeation2014}, the molecules in the adsorption layer contribute a relatively large fraction of the molecules that permeate through. If the adsorbed gas molecules are difficult to permeate through the pore, then they will impede the adsorption and penetration of other gas molecules \cite{SunApplicationnanoporousgraphene2015}. For the situations considered in this work, all the four kinds of foreign gas molecules are unable to permeate through NPG and have stronger adsorption interaction with NPG than H${_2}$. O${_2}$ is more concentrated in the adsorption layer and therefore has a stronger hindering effect towards H${_2}$.

\begin{figure*}[h]
   \centering
   \includegraphics{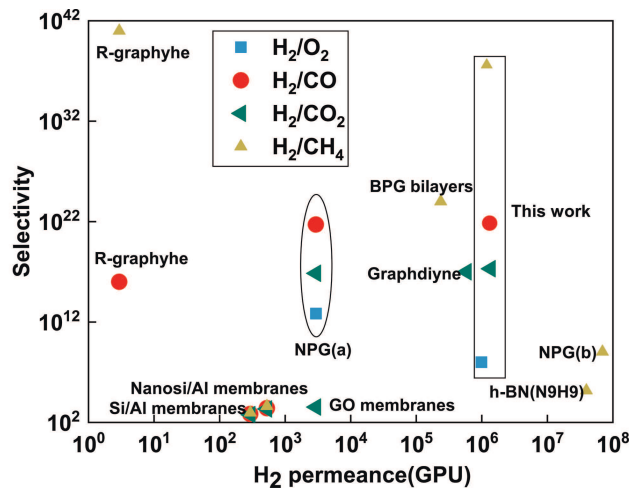}
   \caption{Comparison of the permeability and selectivity of H${_2}$ to O${_2}$, CO${_2}$, CO, and CH${_4}$ in reported studies (Si/Al membranes \cite{GuHydrothermallystablesilica2008}, Nanosi/Al membranes \cite{OyamaTheoryhydrogenpermeability2004}, NPG(a) \cite{BlankenburgPorousGrapheneAtmospheric2010}, NPG(b) \cite{TaoTunableHydrogenSeparation2014}, R-graphyhe \cite{ZhangTunableHydrogenSeparation2012}, Bonded-PG (BPG) bilayers \cite{HuangImprovedpermeabilityselectivity2014a}, GO membranes \cite{LiUltrathinMolecularSievingGraphene2013b} and and Graphdiyne \cite{LiuEnhancedSelectiveHydrogen2020}) and the permeability and selectivity of H${_2}$ to O${_2}$, CO${_2}$, CO, and CH${_4}$ in this study.}
   \label{fig:6}
\end{figure*}

In Figure \ref{fig:6}, we compared the H${_2}$ separation performance of NPG with membrane materials reported previously. It can be found that most of the previously reported H${_2}$ separation membranes have a selectivity below 10$^{24}$ and a permeability below 10$^{6}$ GPU. Some membranes show a high selectivity of 10$^{41}$ but a low permeability. On the other hand, some membranes show a permeability as high as 10$^{8}$ GPU but a selectivity of only 100. In contrast, NPG exhibits superior selectivity while maintaining high permeability of 10$^{6}$ GPU. 

\section{Conclusions}
In summary, we investigated the gas separation properties of NPG using first principles calculations and MD simulations. Firstly, the adsorption behavior of different gases on the NPG surface were calculated using DFT method, and it was found that NPG shows a relatively higher adsorption energy towards O${_2}$. Secondly, the gas permeation energy barriers across NPG was calculated, and it suggests that only H${_2}$ and He show a relatively lower energy barrier which can allow room-temperature permeation. The selectivity of H${_2}$ and He towards other light gases were also estimated. Finally, the permeation behavior of H${_2}$ in gas mixtures was studied by MD simulations, and the simulation results prove that NPG can provide a high H${_2}$ permeability and a superior selectivity simultaneously in various gas combinations. 

\section*{Conflicts of interest}
There are no conflicts to declare.

\section*{Acknowledgements}
The work described in this paper was supported by the National Natural Science Foundation of China (51907159) and Young Talent Recruiting Plans of Xi’an Jiaotong University (DQ6J002). We appreciate computing support from the Xi’an Jiaotong University campus-level public high-performance computing service platform.





 \bibliographystyle{elsarticle-num-names} 
\bibliography{rsc}




\end{sloppypar}
\end{document}